\documentclass{aa}

\usepackage{graphicx}
\usepackage{txfonts}

\usepackage[colorlinks=true,urlcolor=blue,citecolor=red,linkcolor=red,bookmarks=true]{hyperref}

\usepackage{amsmath}
\usepackage{gensymb}

\begin{document}

\title{Extremely young asteroid pair (458271) 2010~UM26 and 2010~RN221} 

\author{D. Vokrouhlick\'y\inst{1},
        P. Fatka\inst{2},
        M. Micheli\inst{3},
        P. Pravec\inst{2},
        and E.J. Christensen\inst{4}}

\titlerunning{Extremely young asteroid pair ...}
\authorrunning{Vokrouhlick\'y et~al.}

\institute{Institute of Astronomy, Charles University, V~Hole\v{s}ovi\v{c}k\'ach 2,
           CZ-180~00 Prague 8, Czech Republic \\ \email{vokrouhl@cesnet.cz}
      \and
           Astronomical Institute, Academy of Sciences of the Czech Republic,
           Fri\v{c}ova 1, Ond\v{r}ejov, CZ 251 65, Czech Republic
      \and
           ESA NEO Coordination Centre, Largo Galileo Galilei 1, I-00044 Frascati,
           Italy
      \and
           Catalina Sky Survey, Lunar and Planetary Laboratory, 1629 E University Blvd,
           Tucson, AZ 85721-0092, USA}

\date{Received: \today ; accepted: ???}

\abstract
{}
{Extremely similar heliocentric orbital elements of the main-belt objects (458271) 2010
 UM26 and 2010 RN221 make them the tightest known pair and promise its very young age. We
 analyzed the conditions of its origin and determined its age.}
{We conducted dedicated observations of (458271) 2010 UM26 and 2010 RN221 in summer 2022
 that resulted in a high-accuracy astrometric set of data. Joining them with the previously
 available observations, we improved the precision of the orbit determination of both asteroids.
 We used numerical simulations backward in time to constrain the origin of this new pair by
 observing orbital convergence in the Cartesian space.}
{Using a large number of possible clone variants of (458271) 2010 UM26 and 2010 RN221 we
 find they all converge in a narrow time interval around March 2003 having extremely tight
 minimum distances ($\leq 1000$ km) and minimum relative velocities ($\leq 3$ cm~s$^{-1}$).
 These conditions require to include mutual gravitational attraction
 of the asteroids constituting the pair for its age determination. Extending our model by this
 effect even improves the convergence results. We find there is more than $55$\% probability
 that the pair formed after the year 2000. However, quasi-satellite captures make the possible
 age uncertainty of this pair prolonged possibly to the 1960s. Still, this is by far the youngest
 known asteroid pair, a prime target for future astronomical observations.}
{}

\keywords{Celestial mechanics -- Minor planets, asteroids: general}

\maketitle


\section{Introduction} \label{intr}
The main belt of asteroids is by no means a static arena. Instead, evolutionary
processes with different timescales permanently sculpt this population into an
ever-changing state. Traditionally, mutual collisions were identified to be the
prime driver of the main belt evolution, leaving behind traces in the form of
asteroid families and occasionally injecting objects into the orbital resonances
(therefore setting them onto a journey to the planet-crossing population). At the
dusk of the last century, the concept of small asteroid migration due to thermal
accelerations was revisited. This process was found to be more important than
collisions in depletion of the main belt, sustaining the transfer of small asteroids
and meteoroids toward the terrestrial-planet zone \citep[e.g.][]{betal2006}. Yet a new
aspect of the main belt asteroid evolution was
found little more than a decade ago when \citet{vn2008} stumbled across a peculiar
population of objects residing on extremely similar heliocentric orbits. Proving the
chances to be so similar by a random fluke is virtually nil, they coined the name
{\em asteroid pairs} for this situation. \citet{vn2008} suggested that the
asteroid pairs may be formed by either of the three mechanisms: (i) in a most
conventional concept, they were assumed to represent a couple of the largest members
in a mini-family (whose other members were not detected yet), (ii) rotational fission
of a parent object, or (iii) recent instability of a binary followed by a split of
its components. \citet{pra2010} used photometric observations of the
primary component of the pairs known of the date to prove that (ii) is the
predominant process that leads to pair formation. This conclusion has
been only strengthened in the most complete study of the asteroid pairs by
\citet{petal2019}. Nevertheless, the relation to the binary asteroids is deep
since it became clear that the asteroid pairs are just failed binaries. Both
pairs and binaries are formed by rotational fission of a precursor asteroid, and
it is basically details of the fission process (such as the component trajectories
immediate after the fission event, the exact shape of the pair components, etc.) that dictates
the outcome \citep[see a review by][]{metal2015}. Discovery of the asteroid pairs thus
complemented a complex picture of a colorful and dynamic life of small asteroids.

The typical age of asteroid pairs, namely the time since their formation, ranges between
several thousands to about a million of years \citep[e.g.,][]{petal2019}. Pairs
with an older age certainly exist too, but their orbits become less similar and
it is difficult to identify them in the unrelated background population of asteroids.
The so far youngest known pairs of main belt asteroids have an age of $\simeq 7$~kyr
\citep[see][]{ziz2016,kyryl2021}. We note the existence of even younger structures, 
such as a pair of near-Earth objects 2019~PR2 and 2019~QR6 \citep{fetal2022}, or 
even fission events caught in (or near) action by direct observations
\citep[e.g.,][]{jetal2015,ye2019,jew2019}. But in these cases, objects of cometary
nature are split by possibly other processes related to their very weak material
strength and/or sub-surface volatile activation.

Here we report the discovery of an extremely young asteroid pair of small main-belt
asteroids (458271) 2010~UM26 and 2010~RN221. The primary
is a little less than a km size object, while the secondary has about $400$
meters in size (assuming $0.2$ geometric albedo and astrometric absolute
magnitudes). Both are located in the central part of the main asteroid
belt, next to the 3/1 mean motion resonance with Jupiter. The similarity
of their orbits is striking even at the standard of other known pairs
\citep[see, e.g.,][]{vn2008}, and it promises their very recent origin. Heading
towards this goal, we first reviewed available astrometry for both objects,
complementing it by recovery on archival frames of the Catalina Sky Survey
(CSS) and remeasuring some of the Canada France Hawaii Telescope (CFHT) data.
We also performed our own targeted
campaign in June and July 2022, obtaining thus very accurate new astrometric
observations. With this dataset available, we performed orbit-determination of
both components in the pair (Sec.~\ref{obs}). Next, we propagated
nominal orbits, and a large number of orbital clones within the uncertainty
limits of the orbit determination, backward in time and monitored their mutual
convergence in Cartesian space. Results from this experiment allowed us to
statistically assess the epoch of separation of the two components in this asteroid
pair (Sec.~\ref{conv}). Finally, we consider further justification for an extremely
young age of this pair and motivate future observations of its components
(Sec.~\ref{disc}). 
\begin{table*}[ht]
\caption{\label{table_elm}	
 Osculating orbital elements and their uncertainty as of epoch MJD $56700.0$ TT.}
\centering
\begin{tabular}{rlccccccc}
\hline \hline
 \multicolumn{2}{c}{Asteroid} & \rule{0pt}{2ex} $a$ & $e$ & $I$ & $\Omega$ & $\omega$
  & $M$ & $H$ \\
 & & [au] & & [deg] & [deg] & [deg] & [deg] & [mag] \\
\hline	
\rule{0pt}{3ex}
458271 & 2010~UM26  & 2.576981320 & 0.326315921 & 3.8602822 & 235.394721 & 119.126545 & 313.470031 & 17.8 \\
       & 2010~RN221 & 2.576985942 & 0.32631528$\phantom{1}$  & 3.860275$\phantom{2}$  & 235.394639 & 119.12674$\phantom{5}$   & 313.467345 & 19.2 \\ [6pt]
 \multicolumn{2}{c}{Uncertainty} & $\delta a$ & $\delta e$ & $\delta I$ & $\delta \Omega$ & $\delta \omega$ & $\delta M$ & $\delta H$ \\
\rule{0pt}{3ex}
458271 & 2010~UM26  & 1.7e-8 & 6.2e-8 & 6.1e-6 & 5.2e-5 & 5.7e-5 & 1.5e-5 & $\sim 0.15$ \\
       & 2010~RN221 & 4.7e-8 & 1.7e-7 & 1.3e-5 & 9.6e-5 & 1.1e-4 & 4.0e-5 & $\sim 0.15$ \\ [2pt]
\hline
\end{tabular}
\tablefoot{Keplerian set of elements used: $a$ semimajor axis, $e$ eccentricity,
 $I$ inclination, $\Omega$ longitude of node, $\omega$ argument of pericenter, and
 $M$ mean anomaly of epoch. The default reference system is that of the heliocentric ecliptic
 J2000. Orbit determination used JPL DE440 ephemerides, and all observations of both
 asteroids from the MPC repository were complemented by our own data (see the Appendix). The absolute magnitude
 values $H$ are provided by the orbit determination solution (thus do not possess
 high photometric quality).}
\end{table*}

\section{Observations and orbit determination} \label{obs}

\noindent{\it Astrometric observations for 458271 and 2010~RN221: archived
and new.-- }As a first step, we overview available astrometric data for 
(458271) 2010 UM26 and 2010 RN221. Both asteroids were discovered during their
favorable 2010 opposition, and subsequently followed-up, mostly with incidental
survey astrometry, during their 2014 and 2018 observation windows. The former
object, being brighter, was also incidentally detected by Pan-STARRS in 2016 and
2020, while fainter than magnitude 22. The latter fainter component was below the
sensitivity of all major surveys during these poorer observability windows.
In addition to this incidental follow-up, precovery observations have been reported
to the MPC for both bodies. The larger one has astrometric coverage from 2005 and
2006, while the fainter has a single listed tracklet from 2005.

With these observations, already present in the MPC database, it is possible to obtain
a good orbit determination for both bodies, sufficient to easily realize the extreme
orbital similarity. However, in order to better understand the dynamics of the pair,
we looked for possible additional prediscovery detections of both bodies. We were able
to locate detections of 2010~RN221 during the 2006 apparition in the image archive of
CSS, and we extracted accurate astrometry from them, including a formal estimate of
the astrometric error bars. In order to ensure the best possible accuracy for the
oldest end of the observational arc, close to the supposed separation time, we also
remeasured the 2005 precovery detections of both asteroids from the original archival
images of CFHT. Our remeasurements were also complemented with properly determined
error bars, not available from the MPC.

Furthermore, both objects became easily observable for astrometric purposes during
the early summer of 2022. We, therefore, decided to complement the dataset with additional
astrometric measurements extending the observed arc to 2022. We observed the field
containing both objects using the $0.8$~m Schmidt telescope at Calar Alto, Spain (MPC
code~Z84), for $5$ nights between May~28, 2022 and July~9, 2022. The larger body was
clearly detectable and measurable on all nights, leading to astrometry with typical
error bars of $\pm 0.2\arcsec$. The smaller body, on the other hand, was only detectable
on three of the nights, due to its faintness and the crowded stellar background. Nevertheless,
astrometry with an accuracy of $\pm 0.3\arcsec$ could be extracted from those nights.
These observations, all obtained using Gaia DR2 as a reference catalog, are presented in
the Appendix, Table~\ref{Astrometry}. Together with the existing data available at the
MPC for the other apparitions, they formed the basis of the orbital analysis presented
in this work.
\smallskip

\noindent{\it Orbit determination: initial data at MJD 56700 epoch.-- }Next,
we used {\tt Find\_Orb} software \citep{gray:2022} to derive new orbits for (458271)
2010~UM26 and 2010~RN221. This solution adopts pre-computed JPL planetary ephemerides
DE440 \citep{park:2021}. Besides barycenters of the eight planetary systems, we also
took into account perturbations caused by Pluto-Charon system and the three largest objects
in the main belt (Ceres, Vesta and Pallas), whose ephemerides were taken from JPL Horizons
System service%
\footnote{Accessible at \url{https://ssd.jpl.nasa.gov/horizons/}}.
{\tt Find\_Orb} provides the best-fit orbital elements, and their full covariance matrix,
for the epoch at the middle of the available observations. While formally the same, we further
propagated their orbits to a common epoch MJD 56700 close to the barycenter of individual solutions
(Table~\ref{table_elm}). This was taken as the initial epoch of our convergence efforts
described in the next Section.

\section{Orbital convergence} \label{conv}
Having set the initial conditions for both components in the (458271) 2010~UM26 and
2010~RN221 pair, we now seek its origin using the past orbital convergence method.
We used {\tt swift\_rmvs4} code, part of the well-tested
N-body package {\tt swift}\footnote{\url{http://www.boulder.swri.edu/~hal/swift.html}},
for our initial tests. For reasons explained below, our final results were obtained
with a more sophisticated N-body integrator known as {\tt SyMBA}
\citep[e.g.,][]{duncan1998}. In both cases, we included perturbations from all planets,
as well as the largest objects in the main belt (Ceres, Pallas, and Vesta). We used a short
timestep of $0.5$~day, and adapted the timespan of the simulation according to its purpose.
We both estimated, and explicitly verified, that thermal accelerations (the Yarkovsky
effect) are not needed to be taken into account thanks to the very young age of this
pair. This is an important simplification, compared to studies of other pairs' origin,
because the thermal accelerations bring along a number of unknown physical parameters
and thus typically blow the age uncertainty.
\begin{figure*}[t]
 \begin{center}
  \begin{tabular}{cc} 
   \includegraphics[width=0.45\textwidth]{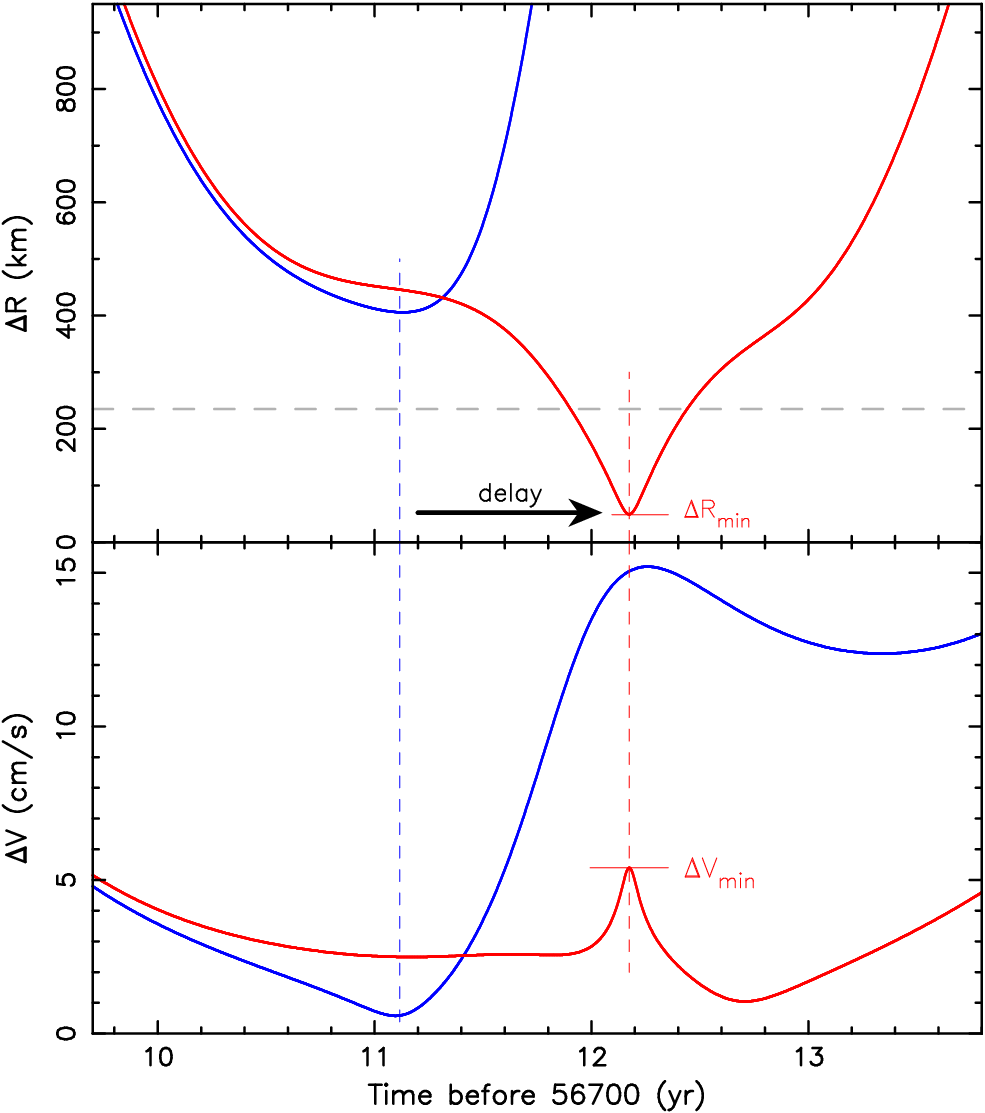} &
   \includegraphics[width=0.45\textwidth]{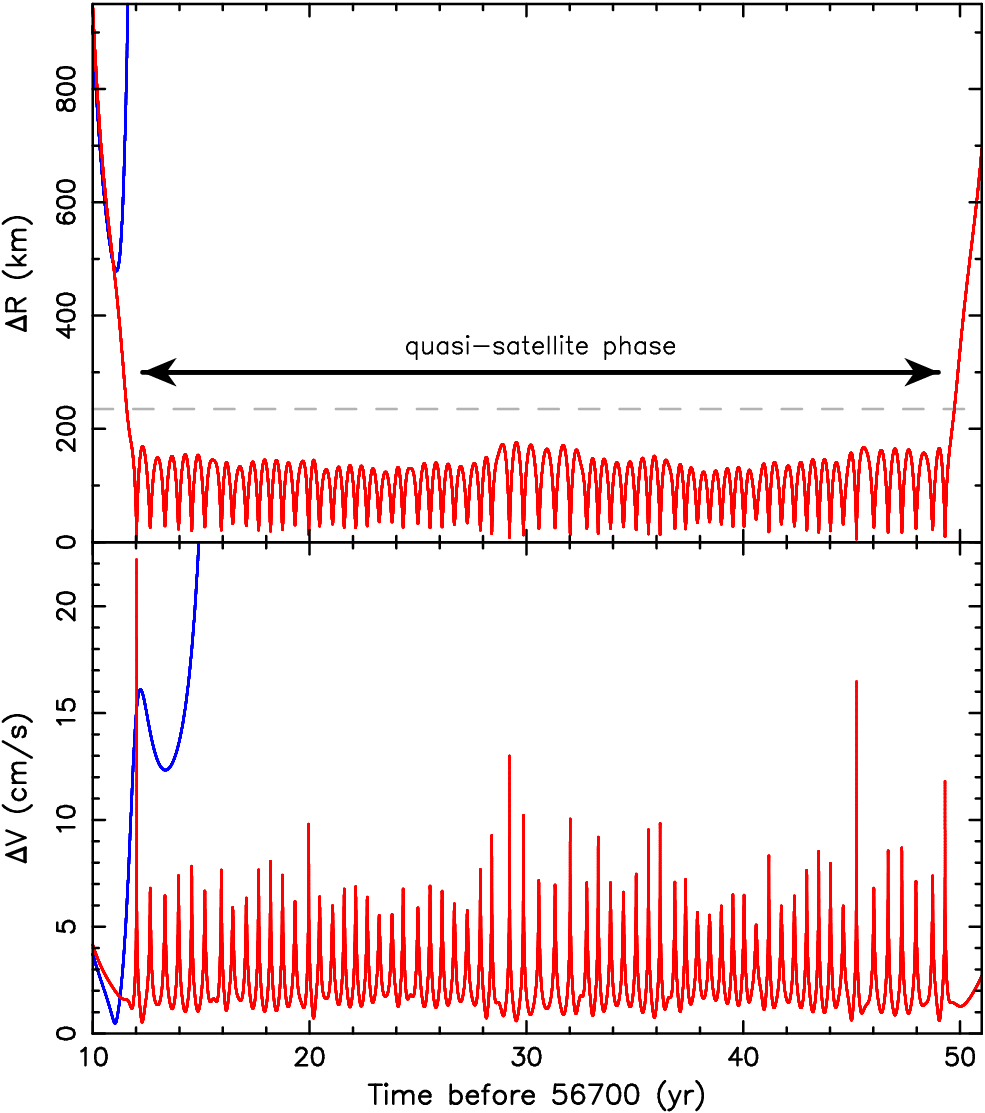} \\
  \end{tabular}
 \end{center}
 \caption{\label{f1}
  Two examples of convergence for (458271) 2010~UM26 and 2010~RN221 orbits.
  The top panels show mutual distance of the two asteroids, bottom panels show
  their relative velocity. The abscissa is the time prior initial conditions
  of the simulation at MJD 56700.0 (Feb~12, 2014). Each simulation was run twice,
  with (red line)
  and without (blue line) the mutual gravitational attraction of the components in
  the pair. The dashed grey line in the top panels indicates the estimated radius
  of the Hill sphere of the parent body. The left panels show a simple close
  encounter at a minimum distance of $\simeq 45$~km distance. The relative orbit
  at the closest encounter is near-to-parabolic flyby diving well into the
  Hill sphere. This effect would have been missed without accounting for the
  mutual gravitational attraction of the asteroids. The right panels show a more complex
  evolution
  when the two asteroids were temporarily captured into a quasi-satellite configuration
  for nearly $40$~years. The orbital eccentricity during the quasi-satellite phase
  is large with apocenter reaching beyond half of the Hill radius.}
\end{figure*}

First, we integrated a large number of clone realizations of the primary ($10000$ clones)
and the secondary ($20000$ clones) over a century timespan backwards in time. Within about
half a year interval centered about March 2003, all possible pairs of clones approach at
a small distance. The distribution of their minimum distances is quasi-Maxwellian with a
peak value of $\simeq 520$~km, a limit representing also $50$\% of cases. Virtually all,
namely $99$\%, clones converge within $\simeq 1000$~km.
These are satisfactorily small values given the estimated radius of Hill sphere of
$R_{\rm Hill}\simeq 230$~km for the parent body of the studied pair. However, the really
distinctive
feature of this recent encounter is the extremely small mutual velocity at which any
combination of clones approached each other: the largest recorded value was $\simeq 4$
cm~s$^{-1}$, but more than $99$\% of them had a mutual encounter velocity of less
than $3$ cm~s$^{-1}$ and more than $68$\% of them had a mutual encounter velocity of less
than $1$ cm~s$^{-1}$.
This is far smaller than the estimated escape velocity from the parent body of the pair,
some $50$ cm~s$^{-1}$. These initial results strongly argue for the very recent split
of (458271) 2010~UM26 and 2010~RN221. At the same time, they also call for a more
detailed modeling effort. This is because when the two asteroids get at the Hill
sphere distance with an extremely small relative velocity, their mutual gravitational
interaction must start to play an important role. Note that this is the first time
when the mutual attraction between the components in the asteroid pair contributes
fundamentally in the convergence modeling: the work of \citet{vn2009} was more of a
curiosity in this respect, and the work of \citet{kyryl2021} was a correct move toward
the concept.  

For the remaining part of this Section we thus switched the integrator to {\tt SyMBA},
which allowed us to account consistently for the mutual gravitational interaction between
the asteroids in the pair. Having no information about the physical parameters of
these bodies, we use their absolute magnitudes, and an assumption of a geometric albedo
of $0.2$, to derive their approximate sizes $\simeq 850$~m and $\simeq 430$~m. We
assumed the bulk density of $2.5$ g~cm$^{-3}$. Because the simulations with {\tt SyMBA}
are more CPU-extensive, we now used a smaller number of clone realizations of the two
asteroids, namely (i) $400$ for 458271, and (ii) $2000$ for 2010~RN221. We performed
$800000$ simulations considering all possible combinations of these clones. We kept the
short timestep of $0.5$~days, and integrated for $70$~yr backwards in time.

As expected, the results from the initial simulation were modified, and substantiated,
by the effects of mutual gravitation attraction of the two asteroids in the pair near
the epoch of their orbital convergence. Two interesting cases are shown in Fig.~\ref{f1}.
The left panel shows a situation when in our previous run the two clones would miss
each other by $\simeq 410$~km at the closest approach (blue curve). However, the gravitational
focusing makes them really approach much closer, to $\simeq 45$~km, performing a nearly
parabolic flyby (red curve). In this situation, the encounter may become delayed by
a year of time. In more regular clone encounters, when even the previous {\tt swift}
simulation made them approach closer along a nearly ``head-on'' approach, the gravitational
focusing advanced the time of the encounter. The right panel of Fig.~\ref{f1} shows a more
extraordinary effect, when the conditions of the pair components approach result in
their capture to a quasi-satellite configuration lasting nearly $40$~years. We found
$\la 10$\% of the clones converging to a minimum distance below $R_{\rm Hill}$ exhibit a
quasi-satellite capture. While small,
their importance is large. This is because we must consider the possibility that 
(458271) 2010~UM26 and 2010~RN221 split at any moment during the quasi-satellite
configuration. Overall, more than $27$\% of all clone combinations result in convergence
well below $R_{\rm Hill}$. Little more than $0.6$\% of cases lead even to a physical collision
of the two asteroids. These numbers increase/decrease with larger/smaller assumed
bulk density or larger/smaller size of the parent object of the pair. In any case, though,
these are large numbers helping to justify a recent split of the components in this pair.
\begin{figure}[t]
 \begin{center} 
 \includegraphics[width=0.47\textwidth]{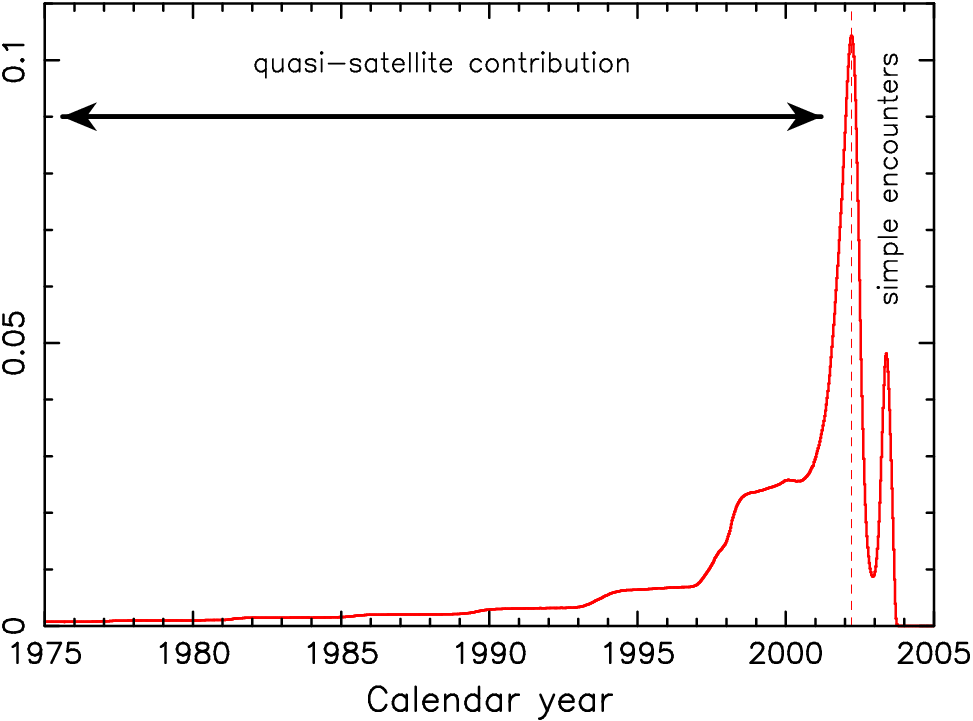}
 \end{center}
 \caption{\label{f2}
  Probability density distribution of the (458271) 2010~UM26 and 2010~RN221 convergence
  epoch. Calendar year at the abscissa, normalization of the ordinate assumes units in days.
  About $8$\% of the solutions are concentrated in a narrow peak centered at about
  $2003.5$. These are simple head-on encounters. Longer-lasting encounters, exemplified in the
  left panel of Fig.~\ref{f1}, and the quasi-satellite solutions, exemplified in the right
  panel of Fig.~\ref{f1}, contribute to the tail extending beyond the peak to the 1960s.}
 \end{figure}

We now combine convergence configurations of all considered couples of clones into
a probability density distribution of the pair origin within the past decades
(considerations about a possibility of larger age are postponed to Sec.~\ref{disc}).
In practice, we record the epoch of all configurations when the clones are (i) closer
to each other than $R_{\rm Hill}$, and (ii) their mutual velocity is smaller than
$50$ cm~s$^{-1}$. We consider their distribution to be a good proxy of the pair
origin. This is because, effects of non-spherical shapes of the components in the
pair or spin-orbit coupling effects during their separation phase may in reality
imply that the two asteroids split and separated from the parent body slightly
earlier or later than just the formal closest approach in our simplified model.
Additionally, the quasi-satellite configurations do not even allow to tell the
moment of the formal closest approach. Again, the complicated shape and spin-orbit
effects may imply that the separation happened at any moment during this phase. Our result
is shown in Fig.~\ref{f2}. Some $8$\% of solutions are concentrated in a peak
at $2003.5\pm 0.25$. These are simple flybys. Before $2003$ the distribution continues
with an extended tail back to the 1960s. This is the contribution of (i) the delayed
long-lasting encounters (as shown on the left panel of Fig.~\ref{f1}), further
convolved with (ii) the quasi-satellite configurations deeper to the past. Despite the
fact that (ii) represents a minority of the solutions, they formally outweigh the flybys in
our approach. This is because they last years, compared to $1$ or $2$ months of the early
flyby configurations. Overall, there is more than $55$\% probability that the pair
formation post-dated year 2000 in our simulation. The exact shape of the probability
distribution may slightly change if statistical dependence on several unknown
parameters (such as bulk density or asteroid size) is included. We postpone such
modeling to the future, when more is known about the components in the pair from
dedicated observation.

\section{Discussion and conclusions} \label{disc}

\noindent{\it Further justification of 458271-2010~RN221
young age.-- }We showed above that the orbits
of (458271) 2010~UM26 and 2010~RN221 converge to a mutual
configuration expected after a fission of their parent body some
$20$ to $60$ years ago. Here we strengthen the case with further
arguments.

First, we show that the configuration at convergence, namely
mutual distance $\leq 1000$~km and relative velocity $\leq
10$ cm~s$^{-1}$ to be conservative, is extremely unlikely to
occur by chance. To that goal, we first determine the intrinsic
collisional probability $p_i$ averaged over all configurations
of asteroid orbits in the main belt. We used the formulation by
\citet{g1982} and more than $91000$ orbits of asteroids with an
absolute magnitude less than $15$ as a proxy to all asteroids.
We obtained $p_i \simeq 3.2\times 10^{-18}$ km$^{-2}$~yr$^{-1}$.
Estimating the number of asteroids larger than a kilometer to
$N_1 \simeq 1.3\times 10^6$, and those larger than $400$~m to
$N_{0.4} \simeq 6\times 10^6$ \citep[e.g.,][]{betal2020},
roughly the sizes of the primary and the secondary components in our 
pair, we now consider how many encounters $N_{\rm events}$
of field asteroids at a distance $R \simeq 10^3$~km (or smaller)
we would expect within the past $T \simeq 50$~yr. We obtain
$N_{\rm events}=p_i N_1 N_{0.4} R^2 T\simeq 1250$ such events.
However, the contributing collisional/encounter configurations
would have a mean relative velocity of $\simeq 5$ km~s$^{-1}$.
Using the formulae in \citet{g1982} we analysed the distribution of
the encounter velocities \citep[see also a detailed discussion of mutual
orbital velocities of asteroids in][namely their Fig.~7 for an idea]{bot1994}.
We found that encounters
with $50$ m~s$^{-1}$ have a likelihood of little more than $10^{-7}$, and
those with even smaller encounter velocities are quite smaller. This shows that
a random encounter of the field (unrelated) asteroids with the indicated
sizes having the required relative velocity $\leq 10$ cm~s$^{-1}$
has a virtually nil probability.

A more difficult task is to show that (458271) 2010~UM26 and 2010~RN221 pair
did not form deeper in the past, probably again by a fission of their
parent asteroid. Indeed, already \citet{vn2008} noted that in orbitally
stable regions of the main belt gently separated bodies may experience
a sequence of repeated encounters with a period equal to their synodic orbital
cycle (see their Fig.~6). Thus a theoretical possibility would be that
(458271) 2010~UM26 and 2010~RN221 separated tens to hundreds of thousands
of years ago, and experienced a close encounter recently after completing a certain
number of synodic cycles. However, there are perturbations, such as encounters
with the largest objects in the belt or differential thermal accelerations,
working against this possibility by generically lifting their encounter
velocity and separating them at larger distances.

In order to estimate the probability of a non-recent origin of (458271) 2010~UM26
and 2010~RN221, we used the methodology introduced in Sec.~4.3 of \citet{ziz2016}.
In particular, we created $10^5$ synthetic pairs by separating a test body from
the nominal orbit of (458271) 2010~UM26 at the reference epoch MJD 56700. We tested
several values of the separation speed from $1$ cm~s$^{-1}$ to $10$ cm~s$^{-1}$,
since these values conformed our convergence solutions in Sec.~\ref{conv}, and
used an isotropic distribution of the initial relative velocity. The orbital
propagation of the separated component, aka 2010 RN221, accounted for the thermal
acceleration (the Yarkovsky effect). The maximum semimajor axis drift has been
estimated using the linearized thermal formulation to $\simeq 5\times 10^{-4}$
au~Myr$^{-1}$ \citep[e.g.,][]{vetal2015}, and a simple implementation of the thermal
acceleration with just a transversal component was used \citep[e.g.,][Sec.~2.1]{far2013}.
To keep the simulation simple, we used {\tt SyMBA} setup with the mutual gravitational
attraction of the components in the pair switched-off. We propagated the orbit
of (458271) 2010~UM26 and the ejected particles for $500$~kyr backward in
time%
\footnote{In a more detailed effort, we would need to consider a large number of
 heliocentric positions and velocities of the primary along its nominal orbit
 in the past. Then create the synthetic clones of the secondary at those epochs, and
 propagate the orbits forward in time to the epoch of the recent encounter. We
 approximate results from such an extensive approach by a simpler integration backward
 in time.}.
We used a short timestep of $1$~day, and at each timestep we monitored the mutual
configuration of (458271) 2010~UM26 and each of the ejected particles in Cartesian
space. We recorded the epochs when these synthetic pairs approached below $1000$~km
and had mutual velocity less than $3$ cm~s$^{-1}$. This is the limit that virtually
{\em all} pairs of clones of (458271) 2010~UM26 and 2010~RN221 reached in the past
two decades (Sec.~\ref{conv}). We started to record these configurations $10$~kyr
through our backward integration, because we estimated by integrations of real
orbits of (458271) 2010~UM26 and 2010~RN221 that the first synodic-cycle approach cannot
occur earlier. We found that the convergence criteria set by the recent approach are
indeed extremely severe and hard to be satisfied on a longer timescale. In the simulation
where the secondaries were separated by the minimum tested velocity of $1$ cm~s$^{-1}$
only about $5$\% of clones ever approached to meet the postulated criteria. Using
higher separation velocities, $3$ cm~s$^{-1}$ and $5$ cm~s$^{-1}$, this fraction even drops
to about $1.4$\% and $0.4$\%, and it becomes negligible for higher separation velocities.
Overall, we thus conclude that the chances that the (458271) 2010~UM26 and 2010~RN221 pair
is older than set in Sec.~\ref{conv} are very low.
\smallskip

\noindent{\it Further analysis and importance of the 458271-2010~RN221
pair.-- }The very recent origin of the (458271) 2010~UM26 and 2010~RN221 pair
motivates further studies. For instance, it would be very interesting to search for
its parent body on archival exposures of large telescopes
in the 1990s or earlier on. Extrapolation of the primary orbit may serve as a
guidance trajectory for this task.

A significant, not yet entirely understood, result from the study of previously known
asteroid pairs is that in all cases with available data the primary and the
secondary were found to rotate about the principal axis of the inertia tensor
\citep[e.g.,][]{petal2019}. This means that (i) either the fission process does
not trigger tumbling of the resulting components, or (ii) internal
energy dissipation in small asteroids is very efficient, leading to damping of
the putative tumbling state in $\la 1-10$~ky. The newly discovered
and extremely young pair (458271) 2010~UM26 and 2010~RN221 may allow
a critical test of these possibilities. We thus urge future photometric
observations of these asteroids with the goal to determine their rotation state.
\citet{petal2019} also found that some $14$\%
of the primaries in known asteroid pairs are binaries themselves. It would
be thus interesting to search for evidence of binarity of the primary
component (i.e., (458271) 2010~UM26). Large aperture telescopes are needed
for both tasks. This is because when bright at opposition, the asteroids of the
pair reported here are unfortunately in the galactic plane in the next decade,
making observations of complex lightcurves of tumbling or binary objects
difficult.
During the oppositions when they are away from the galactic plane, such as in
January~2024, the asteroids are fainter (V magnitude of $21$, or more, even for
the primary).

\begin{acknowledgements}
 This research was supported by the Czech Science Foundation (grant~20-04431S).
 We thank Valerio Carruba for helpful comments on the first version of this
 letter.
 
\end{acknowledgements}


\begin{appendix}

\section{Remeasured and new observations of (458271) 2010~UM26 and 2010~RN221}
In this Section, we provide information about new astrometric data for (458271) 2010~UM26
and 2010~RN221, not previously available at the MPC repository. The sample includes
some data from archival frames, in which we remeasured the positions of the two objects with
greater care, and also our new observations taken in summer 2022 (Sec.~\ref{obs}).
\begin{table*}[ht]
\caption{\label{Astrometry}
  Astrometry of (458271) 2010~UM26 and 2010~RN221 extracted during this work,
  with the corresponding formal error bars in both coordinates. The observing site is
  denoted using its MPC code (third column).}
\centering
\begin{tabular}{rlcccccc}
 \hline \hline
 \multicolumn{2}{c}{Asteroid} & \rule{0pt}{2ex} MPC code & Date (UTC) & $\alpha$ ($\degr$) &
  $\delta$ ($\degr$) & $\Delta\alpha$ ($\arcsec$) & $\Delta\delta$ ($\arcsec$) \\
 \hline
 \rule{0pt}{3ex}
 458271 & 2010 UM26 & 568 & 2005-05-14T09:05:44.0 & 226.661558 & -18.009950 & 0.046 & 0.037 \\
 458271 & 2010 UM26 & 568 & 2005-05-14T09:54:18.0 & 226.652996 & -18.007039 & 0.047 & 0.038 \\
 458271 & 2010 UM26 & 568 & 2005-05-14T10:50:15.7 & 226.643133 & -18.003653 & 0.051 & 0.044 \\
 458271 & 2010 UM26 & Z84 & 2022-05-28T01:01:05.2 & 293.733730 & -17.747669 & 0.25$\phantom{8}$ & 0.25$\phantom{8}$ \\
 458271 & 2010 UM26 & Z84 & 2022-05-30T02:07:33.7 & 293.765217 & -17.632078 & 0.7$\phantom{88}$ & 0.7$\phantom{88}$ \\
 458271 & 2010 UM26 & Z84 & 2022-06-24T22:51:07.3 & 291.382840 & -16.531971 & 0.23$\phantom{8}$ & 0.22$\phantom{8}$ \\
 458271 & 2010 UM26 & Z84 & 2022-06-25T00:12:06.4 & 291.371840 & -16.530500 & 0.17$\phantom{8}$ & 0.17$\phantom{8}$ \\
 458271 & 2010 UM26 & Z84 & 2022-06-25T01:33:01.8 & 291.360718 & -16.529012 & 0.17$\phantom{8}$ & 0.12$\phantom{8}$ \\
 458271 & 2010 UM26 & Z84 & 2022-06-25T02:50:25.7 & 291.350087 & -16.527466 & 0.19$\phantom{8}$ & 0.21$\phantom{8}$ \\
 458271 & 2010 UM26 & Z84 & 2022-07-07T21:22:32.2 & 288.565779 & -16.277297 & 0.2$\phantom{88}$ & 0.2$\phantom{88}$ \\
 458271 & 2010 UM26 & Z84 & 2022-07-08T21:56:06.8 & 288.317946 & -16.265668 & 0.18$\phantom{8}$ & 0.21$\phantom{8}$ \\
 458271 & 2010 UM26 & Z84 & 2022-07-09T00:08:40.2 & 288.294552 & -16.264795 & 0.13$\phantom{8}$ & 0.12$\phantom{8}$ \\
 458271 & 2010 UM26 & Z84 & 2022-07-09T02:21:10.5 & 288.271131 & -16.263635 & 0.27$\phantom{8}$ & 0.21$\phantom{8}$ \\ [6pt]
  & 2010 RN221      & 568 & 2005-05-14T09:05:44.0 & 226.661012 & -18.009844 & 0.068 & 0.063 \\
  & 2010 RN221      & 568 & 2005-05-14T09:54:18.0 & 226.652492 & -18.006931 & 0.077 & 0.072 \\
  & 2010 RN221      & 568 & 2005-05-14T10:50:15.7 & 226.642671 & -18.003558 & 0.09$\phantom{8}$ & 0.086 \\
  & 2010 RN221      & G96 & 2006-12-13T07:52:33.6 & $\phantom{2}$78.082754 & +20.140317 & 0.51$\phantom{8}$ & 0.51$\phantom{8}$ \\
  & 2010 RN221      & G96 & 2006-12-13T08:17:11.0 & $\phantom{2}$78.077912 & +20.139436 & 0.27$\phantom{8}$ & 0.27$\phantom{8}$ \\
  & 2010 RN221      & Z84 & 2022-05-30T02:03:46.2 & 293.754788 & -17.634214 & 0.24$\phantom{8}$ & 0.25$\phantom{8}$ \\
  & 2010 RN221      & Z84 & 2022-06-25T01:13:01.3 & 291.350595 & -16.531642 & 0.34$\phantom{8}$ & 0.34$\phantom{8}$ \\
  & 2010 RN221      & Z84 & 2022-07-09T02:21:10.5 & 288.257447 & -16.266148 & 0.26$\phantom{8}$ & 0.24$\phantom{8}$ \\ [2pt]
\hline
\end{tabular}
\end{table*}

\end{appendix}

\end{document}